\documentclass[12pt]{article}

\usepackage{scicite}

\usepackage{times}

\newenvironment{sciabstract}{%
\begin{quote} \bf}
{\end{quote}}

\newcounter{lastnote}
\newenvironment{scilastnote}{%
\setcounter{lastnote}{\value{enumiv}}%
\addtocounter{lastnote}{+1}%
\begin{list}%
{\arabic{lastnote}.}
{\setlength{\leftmargin}{.22in}}
{\setlength{\labelsep}{.5em}}}
{\end{list}}

\usepackage{graphicx,bm,amsmath,amsfonts,amssymb}
\newcommand{\be}{\begin{equation}}
\newcommand{\ee}{\end{equation}}
\newcommand{\beq}{\begin{eqnarray}}
\newcommand{\eeq}{\end{eqnarray}}
\newcommand{\ba}{\begin{array}}
\newcommand{\ea}{\end{array}}
\newcommand{\bea}{\begin{eqnarray}}
\newcommand{\eea}{\end{eqnarray}}
\newcommand{\ex}[1]{\mbox{e}^{#1}}

\newcommand{\eps}{\epsilon}

\newcommand{\g}{\gamma_{\perp}}

\newcommand{\bma}{\begin{matrix}}
\newcommand{\ema}{\end{matrix}}

\newcommand{\bx}{\bm{x}}
\newcommand{\bxp}{\bm{x}^{\prime}}

\title{Strong Interactions in Multimode Random Lasers}

\author
{Hakan E. T\"ureci$^1$, Li Ge$^2$, Stefan Rotter$^{2,\dagger}$,  A. Douglas Stone$^2$ \\
\\
\normalsize{$^{1}$Institute of Quantum Electronics, ETH Zurich, 8093 Zurich,
   Switzerland}\\
\normalsize{$^{2}$Department of Applied Physics, P. O. Box 208284,} \\
\normalsize{Yale University, New Haven, CT 06520-8284, USA}\\
\\
\normalsize{$^\ast$To whom correspondence should be addressed;
E-mail:  tureci@phys.ethz.ch.}
\\
\normalsize{$^{\dagger}$Presently on leave from Technische Universit\"at Wien,}\\
\normalsize{Wiedner Hauptstra\ss e 8-10/136, A-1040 Vienna, Austria, EU}
}

\date{}

\begin{document}


\baselineskip24pt


\maketitle


\begin{sciabstract}

Unlike conventional lasers, diffusive random lasers (DRLs)  have no
resonator to trap light and no high-Q resonances to support lasing.
Due to this lack of sharp resonances the DRL has presented a
challenge to conventional laser theory.  
We present a theory able to 
treat the DRL rigorously, and provide
results on the lasing spectra, internal fields and output intensities
of DRLs.  Typically DRLs are highly multimode lasers, emitting light at a
number of wavelengths.  We show that the modal interactions through
the gain medium in such lasers are extremely strong and lead to a
uniformly spaced frequency spectrum, in agreement with recent
experimental observations.

\end{sciabstract}

Novel laser systems have emerged recently due to modern
nanofabrication capabilities 
\cite{microcavitylaser,Cao05,PainterLSYODK99}.
The diffusive random 
laser (DRL), perhaps the most challenging of the new
systems, consists of a random
aggregate of particles which scatter light and have gain or are
embedded in a background medium with
gain\cite{Cao05,LawandyBGS94,Mujumdar04,CaoZHSWC99,Cao03,vanderMolenTML07}.
While light scattering in
such a random medium can give rise to Anderson localized, high-Q
resonances \cite{Kumar94,Genack05}, in almost all experiments the
localized regime is not reached and the laser ``cavity" has no
isolated resonances in the absence of gain.
Despite the lack of
sharp resonances, the laser emission from the more recent
DRLs\cite{CaoZHSWC99,Mujumdar04,Cao05} was observed
to have the
essential  properties of conventional lasers: the appearance of
coherent emission with line-narrowing above a series of thresholds,
and uncorrelated photon statistics far from threshold\cite{Cao01}.
Earlier work on random lasers\cite{LawandyBGS94,Cao03} did not 
find
isolated narrow lines and was
interpreted as incoherent lasing, in which there was ``intensity
feedback" but not amplitude feedback.   Later experiments 
\cite{Cao05} and recent numerical
studies\cite{vanneste07} indicate that the lasing involves coherent 
phase-sensitive 
feedback in at least some cases.
Our work shows that standard coherent multimode lasing is possible even
when the linear resonances are much broader than their spacing, 
raising the 
question of what determines the emission frequencies of 
DRLs, since
they are not determined by the position of passive cavity
resonances.
Furthermore, recent experiments on porous GaP DRLs have
shown that the frequencies are rather uniformly spaced and stable
from pulse to pulse, while the intensities vary 
substantially\cite{vanderMolenTML07}.
We show that this is a result of strong non-linear interactions 
between lasing modes combined with extreme leakiness, a regime 
particularly difficult to treat.  In any  multimode laser, the 
different modes compete with one another through  the gain medium in 
a complex manner which depends on the spatial distribution of the 
electric field of each mode. This is particularly severe in the DRL 
where there are many spatially  overlapping modes with similar (very 
short) lifetimes.

The finesse, $f$, of a resonator is defined as the ratio of the
resonance spacing to the
resonance width; standard laser theory only addresses lasers with high finesse
(weakly open) resonators and cannot be applied to the DRL which has
finesse much less than unity.  Hence no analytic results have been
derived relating to 2D or 3D DRLs and realistic numerical simulations
have been limited due to the computational demands. We introduced a
formulation of semiclassical laser theory in terms of biorthogonal
modes, called constant flux (CF) states, which treats lasing media
with any degree of outcoupling and includes the effects of non-linear
modal interactions to all orders\cite{hsc06,TureciSG07}.  We
present analytic and numerical results using this approach
applied to a DRL.

The simplest model for a laser which captures all of the relevant
spatial complexity are the Maxwell-Bloch
equations\cite{haken63,hakenbook}, which are three
coupled non-linear equations for the electric field, the
polarization and the inversion of the gain medium.
For stationary multimode lasing, the modes predicted by
these equations are determined by the time-independent
self-consistent equation\cite{hsc06}
\be
\Psi_\mu (\bf x) =  \frac{i \g}{\g - i (k_\mu - k_a)} \int d\bxp
\frac{D_0(\bx) G(\bx,\bxp;k_\mu)  \Psi_\mu (\bxp)}{\eps(\bxp) (1 + \sum_\nu
\Gamma_\nu |\Psi_\nu (\bxp)|^2)},
\label{eqscint}
\ee
where the electric field is given by
$e(\bm{x},t) = \sum_\mu \Psi_\mu(\bm{x}) \ex{-i\Omega_\mu t}$.
Here the lasing frequencies $\Omega_\mu = c k_\mu$ and the lasing
mode functions $\Psi_\mu(\bm{x})$ are assumed to be unknown (henceforth we
set the speed of light $c=1$ and use the wavevector to denote frequency as well).
In Eq.~\ref{eqscint} 
$k_a$ is the atomic frequency,
$\g$ is the transverse relaxation rate, $\Gamma_\nu = \Gamma (k_\nu)$ 
is the gain
profile evaluated at $k_\nu$,  $D_0(\bx)= D_0 (1 +
d_0(\bx))$ is the pump, which can vary in space, and $\epsilon (\bx)
= n^2(\bx)$ is the dielectric function of the
``resonator".  Electric field and pump strength are dimensionless, 
being measured in
units $e_c = \hbar \sqrt{\g \gamma_{\parallel}}/2g, 
D_{0c}= 4 \pi k_a^2 g^2/\hbar \g$, where 
$\gamma_{\parallel}$ is the 
longitudinal relaxation rate and $g$ is the dipole matrix element of 
the 
gain medium.
Note that each lasing mode $\Psi_\mu$ depends
non-linearly on all of the other lasing modes through the denominator
in Eq.~\ref{eqscint} which represents the ``spatial
hole-burning"\cite{haken63} interaction with the other modes.
Through this
mechanism modes which lase first tend to suppress lasing in other
modes, particularly those with which they are correlated in space.

For simplicity we study a two-dimensional DRL and take the gain
medium to be a uniform disk of radius $R$, which contains randomly
placed nanoparticles with constant index greater than unity.  The
light field in the cavity can be either Transverse Magnetic or Transverse 
Electric polarized
perpendicular to the plane of the disk leading to a scalar equation
for its normal component. The integral in Eq.~\ref{eqscint} is over
the gain region and the kernel  $G(\bx,\bxp;k)$ is the Green function
of the cavity wave equation with purely outgoing boundary
conditions\cite{hsc06}. This represents the steady state
response of the
passive cavity to a monochromatic dipole oscillating with frequency
$k$ at $\bxp$. The non-hermitian boundary conditions on the Green
function lead to a spectral representation $G(\bx,\bxp; k)$ in terms
of dual sets of biorthogonal functions $\varphi_{m}(\bx,k)$ and
$\bar{\varphi}_{m}(\bx,k)$, termed  constant flux (CF) states, with
complex eigenvalues, $k_m$ \cite{hsc06}.

The CF states play the role of the linear cavity resonances 
within our theory and reduce to the quasi-bound
(QB) states within the cavity to a good approximation for a
high Q resonator\cite{bibnote:qbcf}. Importantly, the CF states are
complete within the cavity and generate a conserved photon flux
outside the cavity, unlike the QB states\cite{hsc06}; for the
extremely leaky "cavity" of a DRL the CF and QB states are
significantly different but statistically similar (Fig. 1).

Since the CF basis is complete and conserves flux outside the gain
region, it is an appropriate basis for representing arbitrary lasing
modes $\Psi_\mu (\bx)$ of a DRL. To solve Eq.~\ref{eqscint} we
expand each mode in terms of CF states: $\Psi_\mu (\bx) =
\sum_{m=1}^{N_{CF}} a_m^\mu \varphi_m (\bx)$; substituting this expansion
into Eq.~\ref{eqscint} gives an equation for the complex vector of
coefficients ${\bf a^\mu} = (a_1^\mu, a_2^\mu, \ldots,
a_{N_{CF}}^\mu)$ which completely determines $\Psi_\mu$:
\be
a_m^\mu =
D_0 \sum_n T_{mn}^\mu a_n^\mu.
\label{eqts}
\ee
The non-linear operator $T_{mn}^\mu$ is written out explicitly and
discussed in the Supporting Online Material (SOM) \cite{som}.

This formalism allows us to obtain analytic insight into the question
of what determines the frequencies of the DRL.  In single-mode lasers, each lasing frequency
is a weighted mean of the real part of the cavity resonance frequency
and the atomic frequency\cite{hakenbook}, which for a typical high
finesse system is very close to the cavity frequency with a small
shift (``pull") towards the atomic line. If we denote the
``conventional" lasing frequency by $k_\mu^{(0)}$, we find from
Eq.~\ref{eqts} that for the DRL
\be
k_\mu \approx
k^{(0)}_\mu  + k^{(c)}_\mu
\label{eqkmu}
\ee
where $k^{(c)}_\mu$ is a collective contribution due to all the other
CF states which has no analog in conventional lasers.
In our parameter regime $(k_aR \approx 30)$ both the conventional and
collective terms
are important (although the conventional term is larger) and the
lasing frequencies have no simple relationship to the ``cavity
frequencies".  The collective term is random in sign and
does not always generate a pull towards line center (Fig.~1).
We believe that at larger $k_aR$ the collective term will dominate.


In Fig.~2 we plot the intensities associated with
the lasing modes of
Fig.~1 as a function of pump strength, $D_0$, measured by
the length of the vector of CF coefficients, $I= \sum_{m=1}^{N_{CF}} |
a_m^\mu |^2$.  The behavior is very different from conventional
lasers, showing complex non-monotonic and re-entrant behavior in
contrast to the linear increase found for uniform edge-emitting
lasers \cite{TureciSG07}.  Analysis reveals that the complex
behavior is due to the strong spatial hole burning interactions in
these systems. The inset shows the lasing
frequencies associated with the modes as a function of pump; of the
eight lasing modes in the interval, there are six which form three
pairs nearby in
frequency and their behavior is highly correlated.
Evaluation of the
overlap of
the ${\bf a^\mu }$ vector associated with each pair of modes confirms
that not only their frequencies, but also their
decomposition into CF states are similar.


Equations~\ref{eqscint},\ref{eqts} imply that modes with similar ${\bf a^\mu }$
vectors and similar frequencies will compete strongly because this
leads to a hole-burning denominator which is spatially correlated
with the numerator.  However it is not obvious that frequency
quasi-degeneracy should be associated with spatial correlation for
the DRL. For random lasers with Anderson
localization\cite{Kumar94,Genack05} the CF states would typically be
spatially disjoint,
the $T_{mn} (k)$ operator (cf. Eq.~S3 in \cite{som}) would be
approximately diagonal and there would be no such spatial
correlation. In additional calculations not shown we do find that for 
larger index nanoparticles, which
begin to localize the CF states, 
the modal interactions are reduced. 
But for the DRL, $T_{mn} (k)$ is 
not diagonal and
frequency degeneracy would require an eigenvalue degeneracy in this
complex pseudo-random matrix (see discussion in SOM\cite{som}).
Instead there is eigenvalue repulsion in the complex plane and strong mixing
of eigenvectors, resulting in large spatial overlap of
quasi-degenerate lasing modes and strong hole-burning interaction.
This interaction, in the absence of some special symmetry, tends to
suppress one of the two modes leading to well-spaced lasing
frequencies as found by Ref.~\cite{vanderMolenTML07}.  Hence the
rigid lasing frequency spectrum
could distinguish the DRL from an Anderson localized laser.


This strong interaction of nearly degenerate modes is reflected in a
very large increase in the lasing threshold of the second mode of each pair (see Fig.~2 caption and SOM\cite{som}).
These interaction effects are
strongly non-linear and hence highly sensitive to statistical
fluctuations. To illustrate this in Fig.~3 we
contrast the intensity behavior of Fig.~2, for
which the pump was uniform in space ($d_0(\bf x) = 0$) to a case in
which we have added to the uniform pump a random white noise term
$d_0(\bf x)$ of standard deviation $\pm 30\%$ (normalized to the same total
power).  For this non-uniform pump the third uniform mode (green) now
turns on first. It is thus able to suppress the seventh uniform mode
(purple) over the
entire range of pump powers and acquires almost a factor of three greater
intensity at the same average pump power. The intensities of all the
interacting pairs show similar high sensitivity to pump profile,
while their frequencies remain relatively stable (see inset
Fig.~3).  Exactly such behavior was observed in
shot to shot spectra of DRLs in experiments\cite{vanderMolenTML07}.


Finally, we consider the
spatial variation of the electric field in DRLs (Fig.~4).  The false color
representation of the multi-mode electric field in the laser has a
striking property: it is consistently brighter at the edge of the
disk than at its center, even though the gain is uniform and there
are no special high-Q modes
localized near the edge.  To demonstrate that this effect is not a
statistical fluctuation associated with this particular disorder
configuration we have averaged the behavior of the entire basis set
of CF states over disorder configurations.  The result is a
non-random average growth of intensity towards the boundary.  The
origin of this effect is known from earlier work on Distributed 
Feedback Lasers with
weak reflectivity\cite{kogelnik}; if the single-pass loss is large, then the
single-pass gain must also be large in order to lase, leading to a
visible non-uniformity of the lasing mode, with growth in the
direction of the loss
boundary (on average the radial direction for the DRL). As the
DRL has fractional finesse (which is not achievable in a
one-dimensional geometry)
this effect is much larger in these systems and should be observable.
Note that this effect means
that the electric field fluctuations in DRLs will differ
substantially from the random matrix/quantum chaos fluctuations of
linear cavity modes\cite{qcrmt}, first because each mode is a superposition of
pseudo-random CF states and second because these CF states
themselves are not uniform on average.

The coexistence of gain, non-linear interactions and
overlapping resonances (fractional finesse) makes the DRL
a more complex and richer system
than the widely-studied linear wave-chaotic systems.  It remains to
be seen whether concepts from random matrix theory and semiclassical
quantum mechanics (quantum chaos) will prove fruitful in this
context.  The theory presented here is ``ab initio" in the sense that
it generates all properties of the lasing states from knowledge of
the dielectric function of the host medium and basic parameters of
the gain medium; it should be applicable to any novel laser-cavity
system.

\bibliographystyle{Science}

\begin{thebibliography}{18}
\bibitem{bibnote:qbcf}
{In a high finesse cavity QB and CF states are essentially the same,
each lasing mode is associated with one QB state and the lasing
frequency takes the value $k^{(0)}_\mu$ between the atomic line
center and the cavity frequency $q_1$ determined by the ratio of
$\g/\kappa$ \cite{hsc06} .  }
\end{thebibliography}


\begin{scilastnote}
\item This work was supported by NSF grant DMR-0408636, by the Max Kade 
and W. M. Keck Foundations, and by the Aspen Center
for Physics. We thank Robert Tandy, Manabu Machida, Hui Cao, 
Ad Lagendijk, Patrick
Sebbah, Christian Vanneste, and Diederik Wiersma for discussions.
\end{scilastnote}
\vspace*{2cm}


\noindent 

\begin{figure}
\centering
\includegraphics[width=30.0pc]{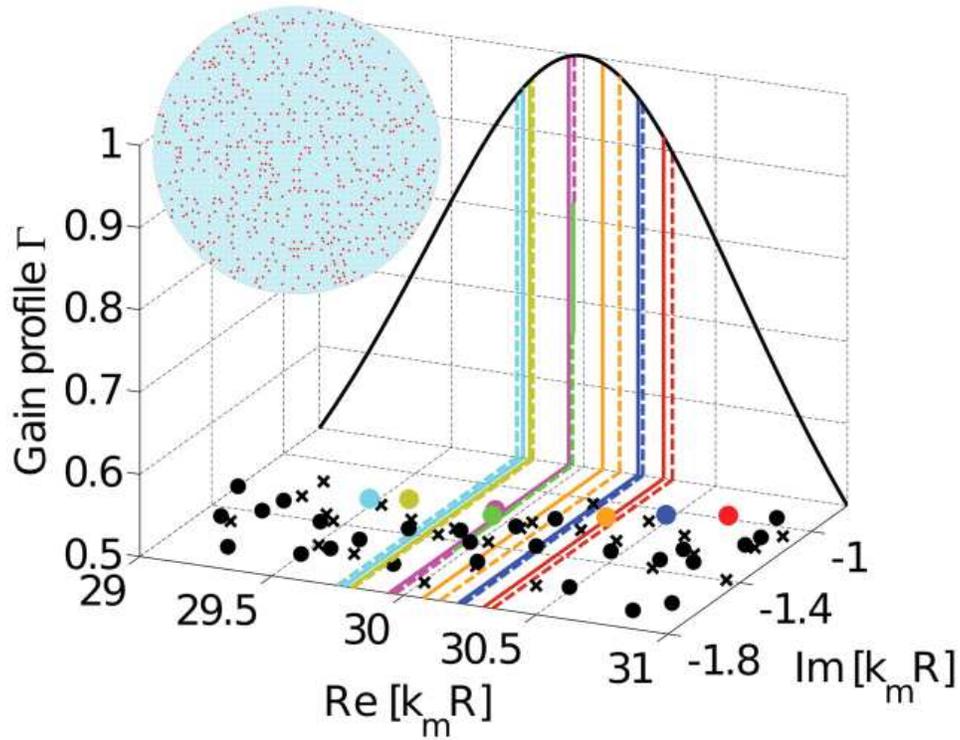}
\caption{Lasing frequencies of a DRL.
Black circles and crosses represent the complex
frequencies of the CF and QB states, respectively; they are
distinct but statistically similar.  As their spacing on the real
axis is much less than their distance from the axis the system has
no isolated linear resonances and the ``cavity" has average finesse less
than unity ($f\approx .05$). Solid colored lines represent the
actual frequencies $k_\mu$ of the lasing modes at pump $D_0/D_{0c} =
123.5035$; dashed lines denote the values $k_\mu^{(0)}$ arising from
the single largest CF state contributing to each mode (the CF
frequency is denoted by the corresponding colored circle).  The thick
black line represents the atomic gain curve $\Gamma(k)$, peaked at the
atomic frequency, $k_a R = 30$.  Because the cavity is very leaky
the lasing frequencies are strongly pulled to the line center in
general, however the collective contribution to the frequency is
random in sign. ({\bf Inset}) Schematic of the configuration of
nanoparticle scatterers in the disk-shaped gain region of the DRL.}

\end{figure}

\noindent
\begin{figure}
\centering
\includegraphics[width=30.0pc]{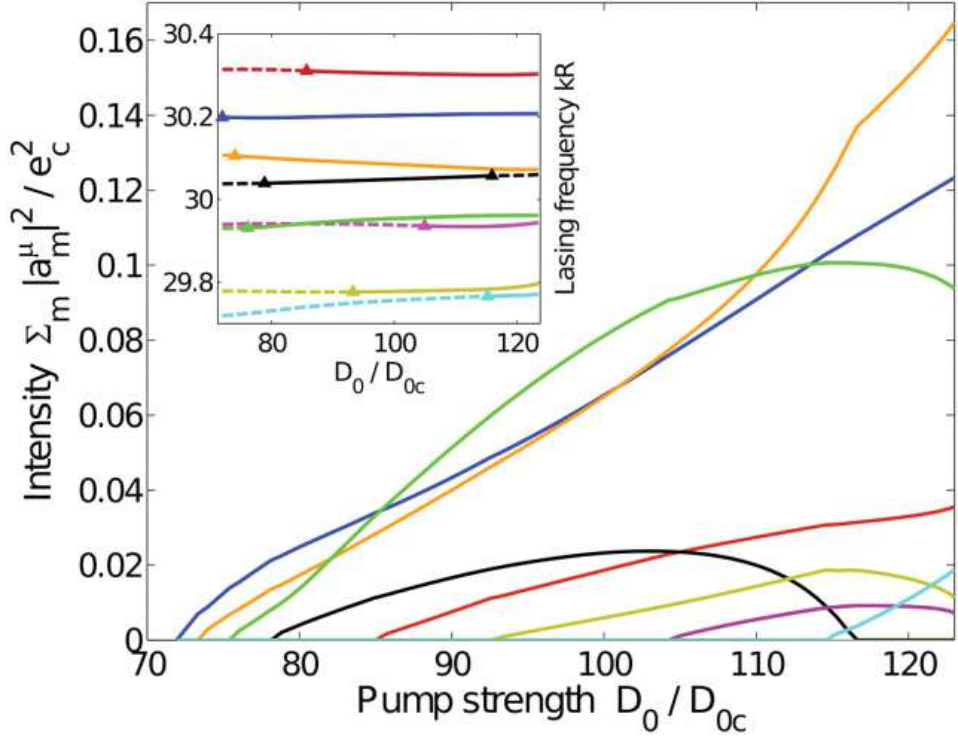}
\caption{Lasing intensities of a DRL. Modal intensities of a DRL vs.
pump $D_0$ for the disorder
configuration of Fig.~1, illustrating complex
non-linear dependence and modal interactions. For example the black
and orange modes approach each other in frequency and interact so
strongly that the black mode is driven to zero, at which point the
orange mode has a kink in its intensity curve. ({\bf Inset}) The
frequencies of each of the eight lasing modes both above (solid) and
below (dashed) threshold vs. $D_0$. Note that frequencies can cross
if one mode is not lasing but not if both are lasing, and modes with
nearby frequencies interact strongly so that their intensity curves
are highly anti-correlated (see discussion in the text).  The
interactions increase the lasing thresholds substantially. For
example the green and purple modes with $k_\mu \approx 29.95$ have
non-interacting thresholds $73.4460,75.6919$ respectively, but the
hole-burning interaction pushes that of the purple mode up to $D_0/D_{0c}
\approx 105$.  In addition there would be sixteen modes lasing by
$D_0/D_{0c} =93$ in the absence of interactions, compared to the six we
find.}
\end{figure}

\noindent
\begin{figure}
\centering
\includegraphics[width=30.0pc]{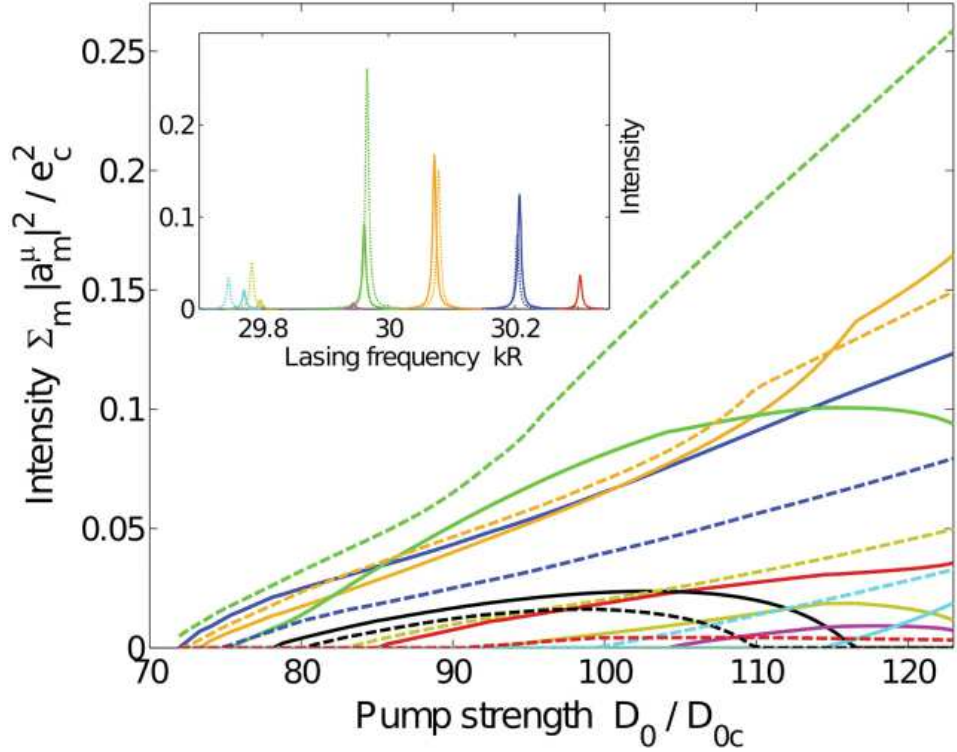}
\caption{Intensity and frequency fluctuations in a DRL. Comparison of
modal intensities of the DRL for the same
disorder configuration in the case of uniform pumping (solid lines)
and partially non-uniform pumping (dashed lines). The main source of
the large intensity fluctuations is the shift in thresholds.  This
has the largest effect for nearly degenerate mode pairs such as the
green and purple modes $(k_\mu \approx 29.95)$. ({\bf Inset}) lasing
spectra at $D_0/D_{0c} = 123.5035$ (lines broadened for visibility). Note
that the black mode doesn't appear at this pump because it has
already been suppressed by the orange mode, and the purple mode only
appears for uniform pumping because it never reaches threshold in
the non-uniform case. The intensities at this pump can fluctuate by
more than a factor of two between the two cases, while the
frequencies fluctuate by just a few percent of their average
spacing.}
\end{figure}

\noindent
\begin{figure}
\centering
\includegraphics[width=30.0pc]{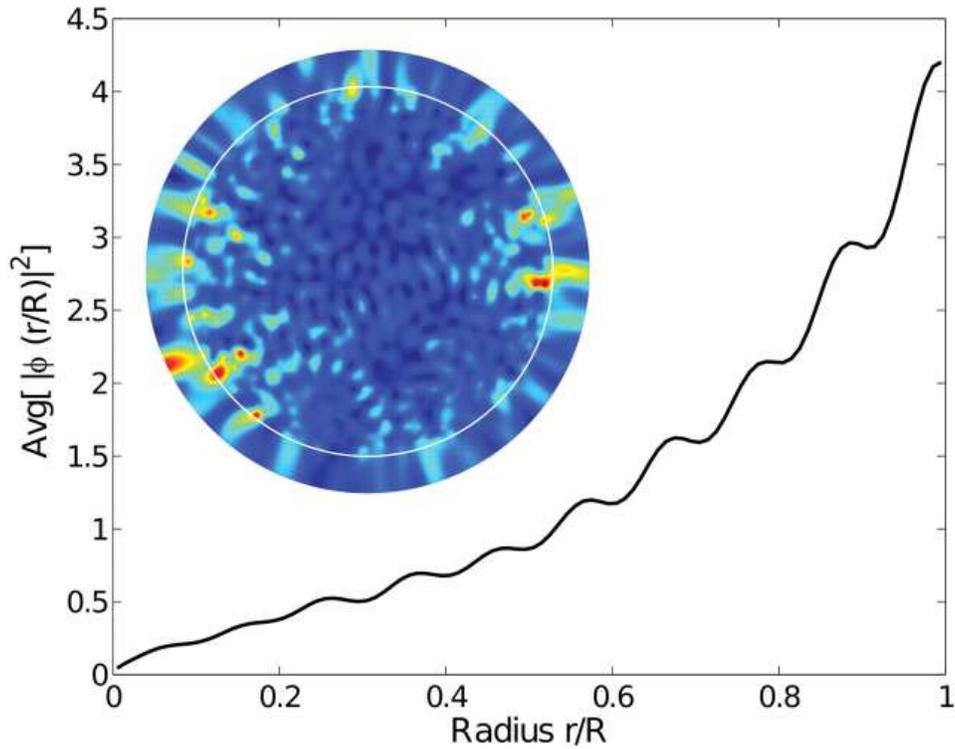}
\caption{Field distribution of a DRL. Radial intensity of CF
states contributing to the lasing modes averaged over 400 disorder
configurations.  There is a large non-random increase of intensity
with radius $r$. ({\bf Inset}) False color plot of
electric field intensity of the
seven lasing modes of the DRL of Fig.~1 at pump $D_0/D_{0c}
= 123.5035$ (white circle is boundary of gain medium). Note brightest
regions appear at the edge of the gain
medium; this is characteristic of low finesse lasers, but is a particularly
large effect in the DRL. }
\end{figure}

\end{document}